\documentclass[prl,showpacs,preprint]{revtex4}
\usepackage{amsmath}
\usepackage{graphicx}
\begin{document}
\title{First-principles study of electron transport through
the single-molecule magnet Mn$_{12}$}
\author{Salvador Barraza-Lopez$^{1}$
\footnote{Present address: School of Physics, Georgia Institute of
Technology, Atlanta, GA 30332},
Kyungwha Park$^1$,
V\'ictor Garc\'ia-Su\'arez$^{2}$, and Jaime Ferrer$^3$}
\affiliation{$^1$Department of Physics, Virginia Polytechnic Institute and State
University. Blacksburg VA, 24061 \\
$^2$Department of Physics, Lancaster University, Lancaster, LA1 4YB, United Kingdom \\
$^3$Departamento de Fisica, Universidad de Oviedo, 33007 Oviedo, Spain }
\begin{abstract}
We examine electron transport through a single-molecule magnet Mn$_{12}$
bridged between Au electrodes using the first-principles method.
We find crucial features
which were inaccessible in model Hamiltonian studies: spin filtering
and a strong dependence of charge distribution on local
environments. The spin filtering remains robust with different molecular
geometries and interfaces, and strong electron correlations, while the charge
distribution over the Mn$_{12}$ strongly depends on them.
We point out a qualitative difference between locally charged and
free-electron charged Mn$_{12}$.
\end{abstract}
\date{\today}
\maketitle



In the past two decades, electron transport through quantum dots has been studied
in single-electron transistors as an effort to manipulate single electrons at a time.
\cite{MEIR91,GOLD98,CRON98}
Semiconducting quantum dots were typically used because of easy manipulation of the
number of electrons inside the dots, by varying gate voltages.

Recently, several experiments \cite{HEER06,JO06,HEND07,VOSS08} on electron transport
through a single-molecule magnet (SMM) Mn$_{12}$ were reported in transistor set-ups
or scanning tunneling microscope (STM) measurements. SMMs differ from magnetic
clusters or quantum dots in the sense that transition metal ions in SMMs are interacting
with each other via super-exchange through ligands, and that there is large magnetic anisotropy
caused by spin-orbit coupling within each SMM.
Thus, the degeneracy in different magnetic states for a given spin multiplet of SMMs
is lifted even in the absence of external magnetic field. The main questions
in these transport studies are whether the electronic and magnetic
properties of SMMs would survive in low-dimensional structures and how the
magnetic degrees of freedom interplay with the electronic degrees of freedom.
The challenges in these types of experiments
are, so far, to maintain stable molecular structures \cite{Fe4}, to determine
orientations of SMMs relative to surfaces, and to characterize interfaces.

Our previous first-principles studies \cite{SALV07,SALV08,SALV08-2} showed
that SMMs are weakly coupled to Au surfaces and electrodes. Thus, transport
through SMMs belongs to a Coulomb blockade regime.
In most theoretical studies on transport through SMMs, one treated
SMMs as quantum dots and relied on many-body model Hamiltonians with unknown
parameter values.\cite{GHKIM04,ROME06,LEUE06,ELST06,GONZ08,MICH08}
However, in contrast to quantum dots, the magnetic properties of SMMs are delicately
balanced by interactions among transition metal ions. Thus, caution needs to be exercised
in interpretation of experimental data to construct effective model
Hamiltonians.

\begin{figure}
\includegraphics[width=13.5 cm, height=7. cm]{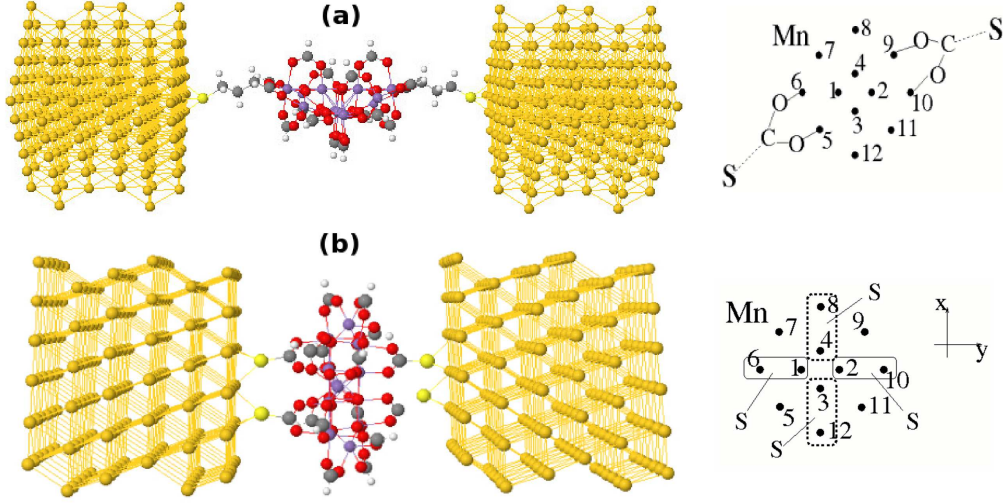}
\caption{(Color online) Scattering region for (a) geometry {\bf 1} consisting
of Mn$_{12}$ attached to Au layers via S atoms and alkane chains
(distance between the electrodes $d$= 25.7~\AA),~and for (b) geometry {\bf 2}
consisting of Mn$_{12}$, four S atoms, and Au layers ($d$=14.5~\AA).~
The transport direction is along the horizontal axis, $z$ axis.
Semi-infinite Au electrodes are considered in calculations (not shown).
On the right hand side the positions of the Mn ions are marked for
each geometry. For geometry {\bf 1} the dashed lines indicate the
alkane chains. For geometry {\bf 2} the solid and dashed
blocks represent the areas where the four S atoms are attached
through bonding to the C atoms.}
\label{fig:geo}
\end{figure}

In this paper, we simulate semi-infinite electrodes and different molecular
geometries and interfaces, and investigate transport properties through
a SMM Mn$_{12}$ bridged between Au electrodes, using the non-equilibrium
Green's function method in conjunction with spin-polarized density-functional
theory (DFT). Our calculations provide crucial microscopic information such as
a spin-filtering effect and a strong dependence of charge distribution over the
Mn$_{12}$ on local environments. This information was unattainable in the theoretical
studies solely based on model Hamiltonians with unknown parameter values, and
could qualitatively change transport properties through SMMs. Mn$_{12}$
molecules used in the transport experiments \cite{HEER06,VOSS08} were bulky
due to large ligands, and so the distances between the Mn$_{12}$ molecule and
the electrodes (or the lengths of linker molecules) for our molecular
geometries are comparable to or much shorter than those in the experiments.
We emphasize the effects of interfaces and molecular geometries on the charge
distribution and the coupling constant between a SMM and electrodes.
To take into account strong electron correlations in
transition metal ions, we included a Hubbard-like $U$ term in our previous
calculations.\cite{SALV08,SALV08-2}

We consider two molecular geometries for a SMM Mn$_{12}$ bridged between Au(111)
electrodes as shown in Figs.~\ref{fig:geo} (a) and (b): (i) geometry {\bf 1} where the
magnetic easy axis of Mn$_{12}$ is perpendicular to the transport direction and
Mn$_{12}$ is attached to the electrodes via alkane chains and S atoms. (ii) geometry
{\bf 2} where the easy axis of Mn$_{12}$ is parallel to the transport direction and
Mn$_{12}$ is attached to the electrodes via four S atoms. 
The linker molecules are used to chemically bind the Mn$_{12}$
to the electrodes, and play a role of energy barriers whose heights depend on
their lengths and types of chemical bonding. 
The Au electrodes are treated
as semi-infinite and the scattering region contains a few Au layers, linker molecules,
and Mn$_{12}$, as shown in Fig.~\ref{fig:geo}. Our calculations are performed using
{\tt SMEAGOL} \cite{SMEAGOL,FERN06}, a quantum transport code interfaced with
DFT {\tt SIESTA} program \cite{SIESTA}.
Generalized-gradient approximation (GGA) \cite{PERD96} is used for exchange-correlation
potential in spin-polarized DFT formalism.
When spin-orbit coupling is included self-consistently for an isolated neutral Mn$_{12}$
molecule, magnetic anisotropy barrier is computed to be 65.3~K using {\tt VASP}\cite{VASP}
and 66.4~K using {\tt SIESTA},\cite{FERN06} which agrees well with experimental data.\cite{BARR97}
Gate voltages and interactions with phonons are not considered in this study.
Further details of the method and assumptions for this work and brief
discussion on the transport for geometry {\bf 1} were presented in Ref.\cite{SALV08-2}.


\begin{figure}
\includegraphics[width=7.8cm, height=5.4cm]{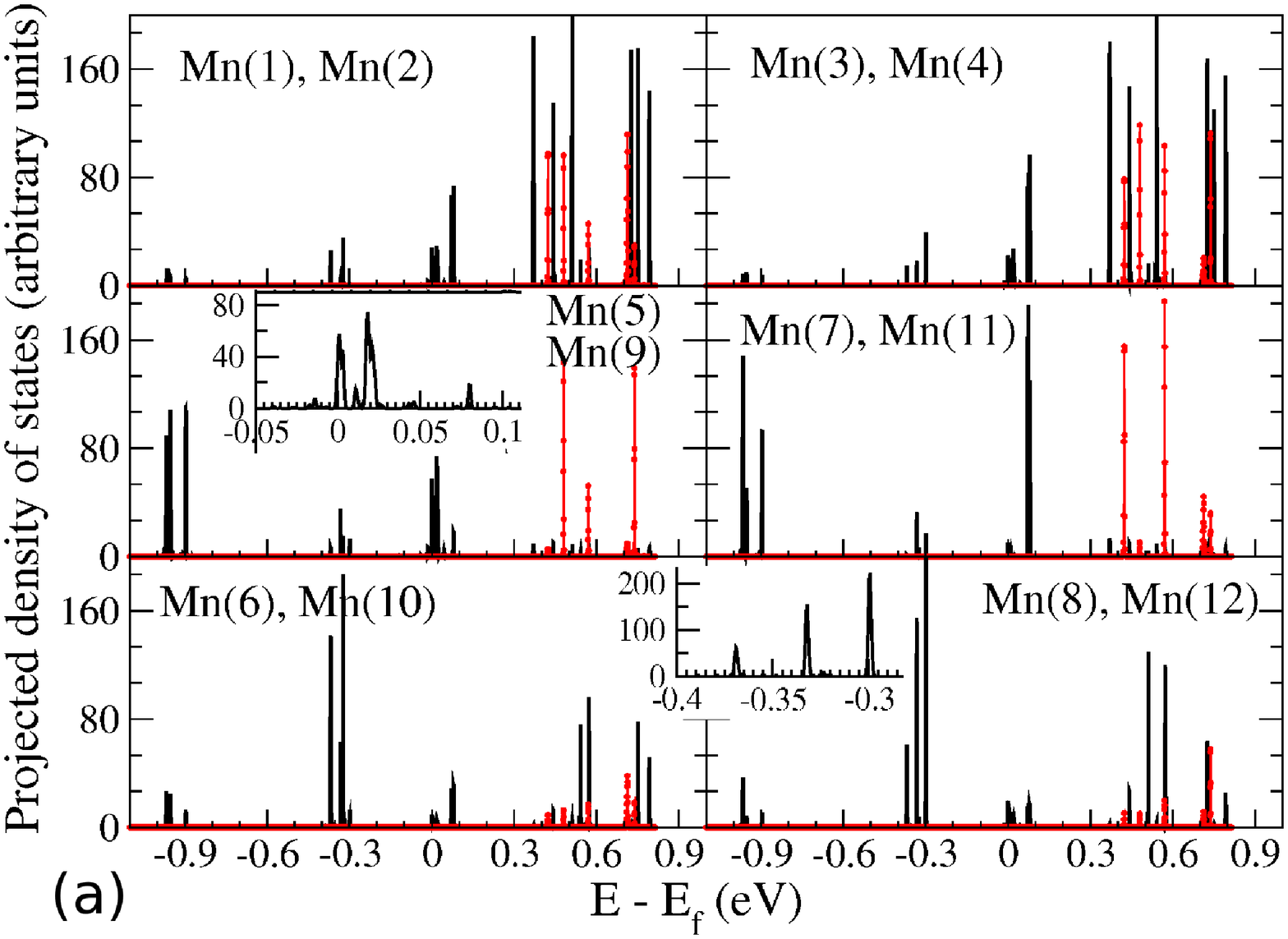}
\hspace{0.5truecm}
\includegraphics[width=7.8cm, height=5.4cm]{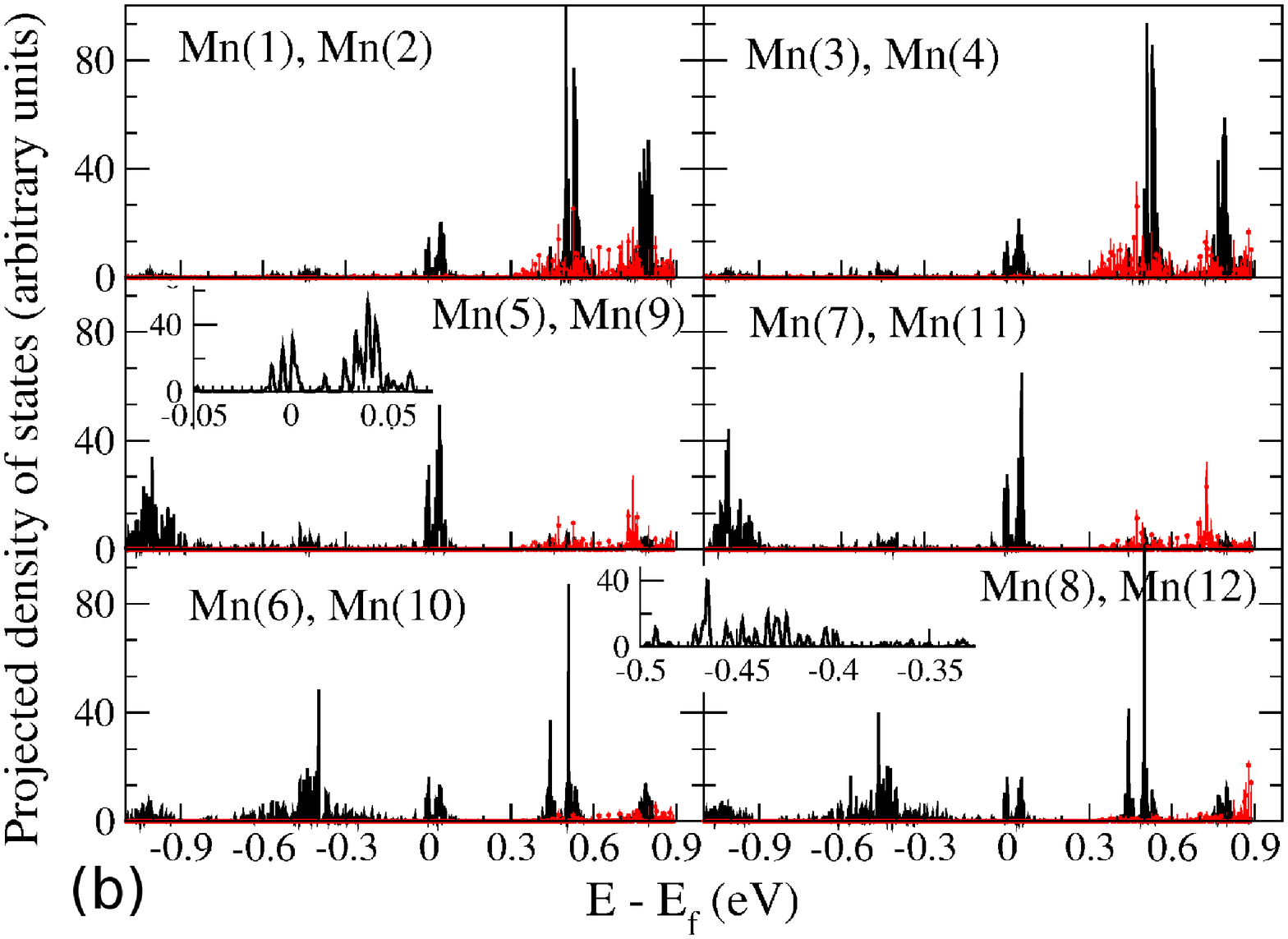}
\caption{(Color online) Spin-polarized density of states (DOS) projected onto Mn $d$ orbitals
(a) for geometry {\bf 1} and (b) for geometry {\bf 2}: majority spin (black),
minority spin (red with symbol). Refer to Fig.~\ref{fig:geo} for numbering of the Mn ions.
Insets: zoom-in of the majority DOS onto Mn(5), Mn(9), Mn(8), and Mn(12).}
\label{fig:PDOS-2}
\end{figure}

The spin-polarized densities of states (DOS) of the scattering region projected onto all
Mn $d$ orbitals for geometries {\bf 1} and {\bf 2} are shown relative to the
Fermi level, $E_f$, in Figs.~\ref{fig:PDOS-2}(a) and (b). In the DOS for both geometries
a bin size of 0.5~meV and Gaussian broadening of 1~meV are used. The minority-spin
DOS become negligible in the energy window (-1.6, 0.42~eV) and (-1.5, 0.3~eV)
relative to $E_f$ for geometries {\bf 1} and {\bf 2}, respectively. Thus, for both geometries
only the majority-spin orbitals contribute to the densities near $E_f$. Further discussion of the
DOS for geometry {\bf 1} is followed by that for geometry {\bf 2}.

For geometry {\bf 1} the fourfold symmetry of Mn$_{12}$ is broken due to the linker molecules
so that the degeneracy in the molecular orbitals is lifted. As shown in Fig.~\ref{fig:PDOS-2}(a), 
the projected densities of states (PDOS) for Mn(5) and Mn(9) completely differ from those for Mn(7)
and Mn(11), although the four Mn sites are equivalent according to the fourfold symmetry.
The individual molecular orbitals are clearly identifiable due to the larger distance between
the Mn$_{12}$ and the electrodes (or longer linker molecules) compared to those in geometry {\bf 2}. 
The coupling (or broadening) of the orbitals to the electrodes near
$E_f$ is of the order of 1~meV or less. The lowest unoccupied molecular orbital
(LUMO) is located slightly above $E_f$ and the highest occupied molecular orbital (HOMO)
is placed 0.29~eV below $E_f$ [Fig.~\ref{fig:PDOS-2}(a)]. The LUMO is mainly from two of the Mn
ions in the outer ring, Mn(5) and Mn(9), while the HOMO is from Mn(8) and Mn(12).

For geometry {\bf 2} the shorter distance between the Mn$_{12}$ and the electrodes allows 
the molecular orbitals to substantially broaden, which makes the individual orbitals 
unidentifiable [Compare the insets of Figs.~\ref{fig:PDOS-2} (a) and (b)]. That distance 
for geometry {\bf 2} is about a half of that for geometry {\bf 1}, but the increase in the coupling
constant for geometry {\bf 2} is much greater than a factor of 2 due to the exponential
decay of the coupling with the distance for a given chemical bonding. The group of peaks near
$E_f$ is formed by broadening of the LUMO, which arises from coupling of all of the Mn ions
to the electrodes, in contrast to the case of geometry {\bf 1}.
The HOMO broadens due to coupling of Mn(6), Mn(8), Mn(10), and Mn(12) to the electrodes.
In geometry {\bf 2}
the fourfold symmetry of an isolated Mn$_{12}$ is, to some extent, preserved, because the
linker molecules are attached in a fourfold symmetric fashion.
The coupling constant between the Mn$_{12}$ and electrodes
near $E_f$ is of the order of 10~meV [insets of Fig.~\ref{fig:PDOS-2}(b)]. 


\begin{figure}
\includegraphics[width=7.8cm, height=5.3cm]{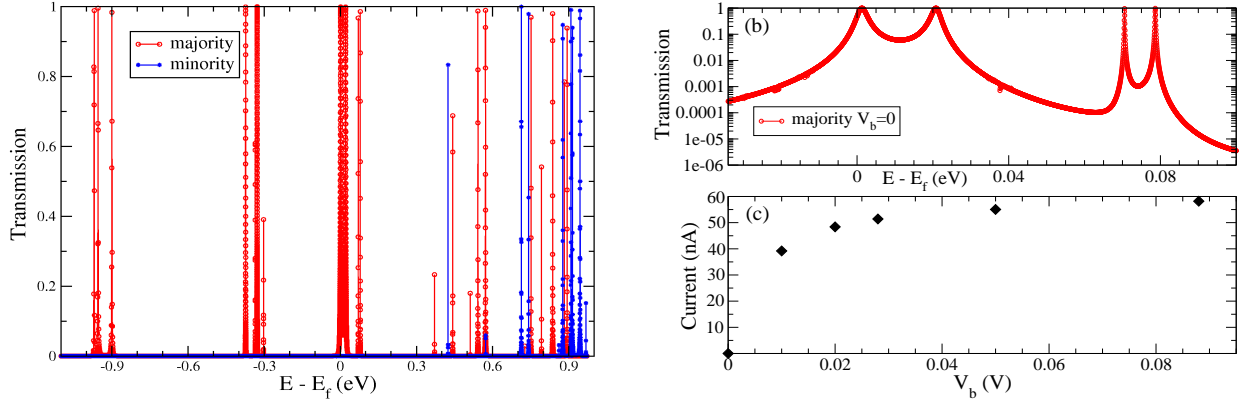}
\hspace{0.5truecm}
\includegraphics[width=7.8cm, height=5.3cm]{Extra-Fig3-Apr27-v07.eps}
\caption{(Color online) (a) Spin-polarized transmission coefficient at zero bias,
(b) zoom-in of the majority-spin transmission at zero bias [(a)], 
and (c) computed current $I$ vs bias voltage $V_b$ for geometry {\bf 1}.} 
\label{fig:TRC}
\end{figure}

\begin{figure}
\includegraphics[width=7.8cm, height=4.9cm]{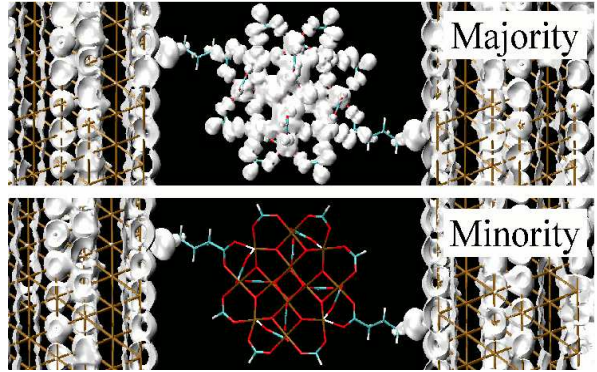}
\caption{(Color online) Majority-spin and minority-spin density of states integrated
between -0.23 and 0.06 eV relative to the Fermi level with isosurface
criterion of 2~$e$/nm$^3$ for geometry {\bf 1}.}
\label{fig:PDOS-3}
\end{figure}

We compute a spin-polarized transmission coefficient $T(E)$ for geometry {\bf 1} and
our result at zero bias is shown in Fig.~\ref{fig:TRC}(a). The majority-spin LUMO is
responsible for the resonant tunneling near $E_f$. The widths of the $T(E)$ peaks,
in general, depend on broadening of the orbitals, phonon populations, and defects.
In our case, since we did not include defects or interactions with phonons, the
widths depend on the broadening only. The weak coupling leads to the widths of
the $T(E)$ peaks ranging from 0.01 to 1~meV [Fig.~\ref{fig:TRC}(b)].
The minority-spin contribution
to $T(E)$ appears only 0.42~eV above $E_f$ and 1.6~eV below $E_f$. This agrees
with the locations of the orbitals in the PDOS [Fig.~\ref{fig:PDOS-2}(a)].
For geometry {\bf 1} there is a one-to-one mapping between the $T(E)$ peaks
and the orbitals. The spatially resolved density of states integrated over (-0.23, 0.06~eV)
relative to $E_f$ (Fig.~\ref{fig:PDOS-3}) clearly corroborates that minority-spin
electrons cannot tunnel through the Mn$_{12}$ at low bias voltages.
Using the same analogy, we expect that for geometry {\bf 2} only the majority-spin
orbitals would contribute to $T(E)$ near $E_f$, and that the $T(E)$ peaks would be 
wider due to the broadening of the orbitals. As shown in Fig.~\ref{fig:TRC}(c),
for geometry {\bf 1} our computed current as a function of bias voltage $V_b$ is of
the order of tens of nA when $V_b < 0.1$~V. 
For a given positive $V_b$, the chemical potential of the left (right) electrode
increases (decreases) by $eV_b/2$. We find that for 
$0 < V_b < 0.1$~V, $T(E)$ is shited upward compared to the zero-bias $T(E)$ without
changing the main features. This shift is caused by the polarization of the Mn$_{12}$
with $V_b$ and the amount of the shift is proportional to $V_b$. Then the area of
$T(E)$ integrated over $E$ starts to saturate
above 0.02~V, which results in the saturation in the current-voltage dependence
[Fig.~\ref{fig:TRC}(c)].

To compare with experiment, additional electron correlations within the Mn ions that
are absent in standard DFT must be considered. Beyond DFT, the $U$ term was included
in our calculation of the electronic
structure of an isolated Mn$_{12}$ molecule \cite{SALV08} using {\tt VASP}.
It was found that the majority-spin HOMO-LUMO gap increased to 1.1~eV and that the
majority-spin LUMO was shifted upward by 0.12~eV. The minority-spin LUMO was, however,
still 0.12~eV above the majority-spin LUMO. Thus, when an extra electron is added to
the Mn$_{12}$, the majority-spin orbitals are still well separated from the minority-spin
orbitals. Consequently, only majority-spin electrons can be tunneled through the Mn$_{12}$
at low bias voltages (below 0.5 eV). Thus, we conclude that the spin-filtering
effect remains robust with different molecular geometries and interfaces, and strong
electron correlations.

\begin{figure}
\includegraphics[width=6.cm, height=4.cm]{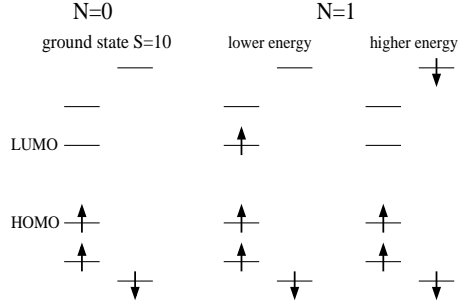}
\caption{Spin non-degenerate molecular orbitals of neutral Mn$_{12}$
($N=0$) in the ground state ($S=10$) and of singly-charged Mn$_{12}$
($N=1$) in the ground state and an excited state, obtained from DFT.
In each state, four majority-spin (two minority-spin) orbitals are shown
on the left (right). }
\label{fig:orbitals}
\end{figure}

Henceforth, we discuss a subtle but important issue in the transport through SMM
Mn$_{12}$. The computed charging energy of an isolated Mn$_{12}$ is 3.8~eV and thus
the Mn$_{12}$ can be only singly charged at low bias voltages.
Due to the nature of less-than-half-filled $d$ orbitals, the Mn $d$ orbitals
in the Mn$_{12}$ are not spin-degenerate.
If an extra electron is added to Mn$_{12}$, the ground state would be achieved
when the added electron has majority spin (Fig.~\ref{fig:orbitals}).
This is the origin of the spin-filtering
effect in the transport through Mn$_{12}$. However, this argument must not be
straightforwardly interpreted that the ground-state spin of [Mn$_{12}$]$^{1-}$, is
$S=10+1/2=21/2$. This is true only if the ground-state spin of
[Mn$_{12}$]$^{1-}$ has a collinear configuration or if the extra charge is
distributed over several Mn sites instead of being localized at one Mn site.

When conduction electrons are added to the Mn$_{12}$ from the electrodes, our
DFT calculations with collinear spin configurations support that the electrons
will be distributed over more than one Mn site. The specific distribution of the
electrons over the Mn sites depends on the way the Mn$_{12}$ molecule is attached
to the electrodes.
For geometry {\bf 1} the electrons will be mainly distributed over the Mn(5) and Mn(9)
sites, while for geometry {\bf 2} they will be distributed over all of the Mn sites
[Figs.~\ref{fig:PDOS-2}(a) and (b)].
Our noncollinear DFT calculations on [Mn$_{12}$]$^{1-}$ (using {\tt SIESTA}) reveal
that the collinear spin configuration with $S=21/2$ has the lowest energy.

Some transport studies based on model Hamiltonians,\cite{HEER06,ROME06,GONZ08}
assumed that a singly charged Mn$_{12}$ molecule has the ground-state spin of $S=19/2$,
by referring to experiments \cite{EPPL95,BASL05} performed on locally charged Mn$_{12}$
molecules. 
In these experiments, extra electrons were added to the magnetic core of the Mn$_{12}$,
by inserting cations (or electron donors) close to one or two of the Mn ions. Since a bulk form
of Mn$_{12}$ molecules is an insulator, the added electrons would be localized
to the Mn sites closest to the cations. The experiments showed that the total
spin for [Mn$_{12}$]$^{-1}$ was $S=19/2$, and that one of the Mn ions
in the outer ring changed its valence from $3+$ ($S=2$) to $2+$ ($S=5/2$).
In contrast to typical quantum dots, the total spin of Mn$_{12}$ is
maintained through delicate balance among interactions between the different Mn ions.
We perform noncollinear calculations on Mn$_{11}$Fe (one of the Mn ions in the outer ring
is replaced by Fe in the Mn$_{12}$ geometry) in order to mimic a singly locally charged
Mn$_{12}$. Using $U_{\rm Mn}$=4~eV \cite{SALV08} and $U_{\rm Fe}$=6~eV in {\tt VASP},
we find that the collinear spin configuration with $2S=19$ has 100~meV higher energy than
the collinear spin configuration with $2S=21$. The latter has 5.8~meV higher energy
than the noncollinear spin configuration ($2S=18.71$) in which the magnetic moment
vector of the Fe$^{3+}$ ($S=5/2$) ion is tilted by 62$^{\circ}$ from the moment vectors
of the eleven Mn ions mostly aligned along with the $z$ axis.
Additional noncollinear
calculations on the Mn$_{11}$Fe with $U=0$ also supports that the collinear spin
configuration with $2S=21$ is not the ground state for the singly locally charged Mn$_{12}$.
Thus, the spin-filtering effect and single-electron picture (Fig.~\ref{fig:orbitals})
are compatible with the experimental findings \cite{EPPL95,BASL05}.

In summary, we have investigated transport properties through a Mn$_{12}$ molecule
bridged between Au(111) electrodes using the non-equilibrium Green's function method and
spin-polarized DFT.
We found that the Mn$_{12}$ functioned as a spin filter in a low bias regime (below 0.5~eV).
The spin-filter effect persisted with different molecular geometries and interfaces and
strong electron correlations, while the distribution of conduction electrons over the
Mn$_{12}$ strongly depended on them.
Conduction electrons from the electrodes would be distributed over
several Mn sites rather than being localized at one Mn site.
There is no contradiction between the spin-filtering effect and the experimental
observation on $S=19/2$ for the singly locally charged Mn$_{12}$ molecules.






K.P. was supported by NSF DMR-0804665, the Jeffress Memorial Trust Funds,
and NCSA under DMR060009N. J.F. was supported by MEC FIS2006-12117.
The authors are grateful to W. Wernsdorfer for discussions.

\end{document}